\documentclass[prl, aps,  twocolumn,superscriptaddress,
showpacs]{revtex4-2}

\usepackage[colorlinks,urlcolor=blue,citecolor=blue,linkcolor=blue]{hyperref}

\usepackage{physics}
\usepackage[english]{babel} 
\usepackage[english]{layout}
\usepackage{amsmath,amsfonts,amssymb}
\usepackage{graphicx}
\usepackage{float}
\usepackage{color}
\usepackage{braket}
\usepackage{version}
\usepackage{array,multirow,makecell}
\usepackage{afterpage}
\usepackage[toc,page]{appendix} 
\usepackage{bbold}

\usepackage{soul}

\renewcommand{\k}{{\bf k}}

\newcommand{\q}{{\bf q}}

\newcommand{\sch}{Schr{\"o}dinger }

\renewcommand{\k}{{\bf k}}

\newcommand{\beq}{\begin{equation}}
\newcommand{\eeq}{\end{equation}}

\usepackage{xcolor}
\newcommand{\new}[1]{{\color{black}#1}}

\begin{document}

\title{Scattering resonances and pairing in a Rabi-coupled Fermi gas}

\author{Olivier Bleu}
\thanks{These two authors contributed equally to this work.}
\affiliation{School of Physics and Astronomy, Monash University, Victoria 3800, Australia}

\author{Brendan C. Mulkerin}
\thanks{These two authors contributed equally to this work.}
\affiliation{School of Physics and Astronomy, Monash University, Victoria 3800, Australia}

\author{Cesar R. Cabrera}
\affiliation{Institute for Quantum Physics, Universität Hamburg, Luruper Chaussee 149, 22761 Hamburg, Germany}
\affiliation{The Hamburg Centre for Ultrafast Imaging, Universität Hamburg, Luruper Chaussee 149, 22761 Hamburg, Germany}

\author{Jesper Levinsen}
\affiliation{School of Physics and Astronomy, Monash University, Victoria 3800, Australia}

\author{Meera M. Parish}
\affiliation{School of Physics and Astronomy, Monash University, Victoria 3800, Australia}

\date{\today}

\begin{abstract}
We investigate the possibility of using a Rabi drive to tune the interactions in an atomic Fermi gas. Specifically, we consider the scenario where two fermion species (spins) are Rabi coupled and interacting with a third uncoupled species. Using an exact calculation within a minimal low-energy model, we derive analytical expressions for the effective scattering length and effective range that characterize the collisions between a Rabi-dressed atom and an atom from the third species. In particular, we find that new scattering resonances emerge in the Rabi-coupled system, which we demonstrate are linked to the existence of hybrid two-body bound states. Furthermore, we show via a generalized Thouless criterion  that the scattering properties have a direct impact on the superfluid transitions in the Rabi-coupled Fermi gas. The presence of Rabi-induced resonances thus has implications for the investigation of many-body physics with driven atomic gases.
\end{abstract}

\maketitle

The Feshbach resonance is a cornerstone of ultracold atomic gas experiments~\cite{Chin2010}. By tuning a molecular two-body bound state into resonance using, e.g., an external magnetic field, the interatomic collisions can be precisely controlled, thus providing an essential tool for simulating correlated quantum phenomena~\cite{Bloch2012}. This has permitted the experimental realization of a range of many-body systems, including paired-fermion superfluids~\cite{Regal2004,Zwierlein2004,Bourdel2004,Chin2004} and polaron quasiparticles~\cite{Schirotzek2009,Nascimbene2009,Hu2016,Jorgensen2016}, as well as exotic few-body states such as Efimov trimers~\cite{Kraemer2006,Lompe2010}. 

Another important component of the cold-atom toolbox is the ability to manipulate atoms with electromagnetic fields. 
In particular, the coupling of two atomic species using radio, microwave or optical frequency pulses has provided a powerful spectroscopic probe of the state of an atomic gas, both in the frequency and the time domains~\cite{punk2007,Cetina2016,Vale2021}. Even in the case of a continuous drive, which goes beyond the linear response regime, the resulting Rabi oscillations can yield information about quantum behavior such as the superfluid order parameter in a Bose Einstein condensate (BEC)~\cite{Matthews1999} or Landau quasiparticle properties in a Fermi gas~\cite{kohstall2012,Koschorreck2012,Scazza2017,Oppong2019,Adlong2020}. Most recently, there is the exciting prospect of using a Rabi drive to fundamentally modify the quantum gas itself, a situation which has already been achieved in two-component %
Bose mixtures~\cite{Zibold2010,Nicklas2011,Lavoine2021,Hammond2022,Sanz2022,Cominotti2023,Zenesini2024} and for a highly spin-imbalanced Fermi gas~\cite{Vivanco2025}.

\begin{figure}[tbp] 
\centering
    \includegraphics[width=1\linewidth]{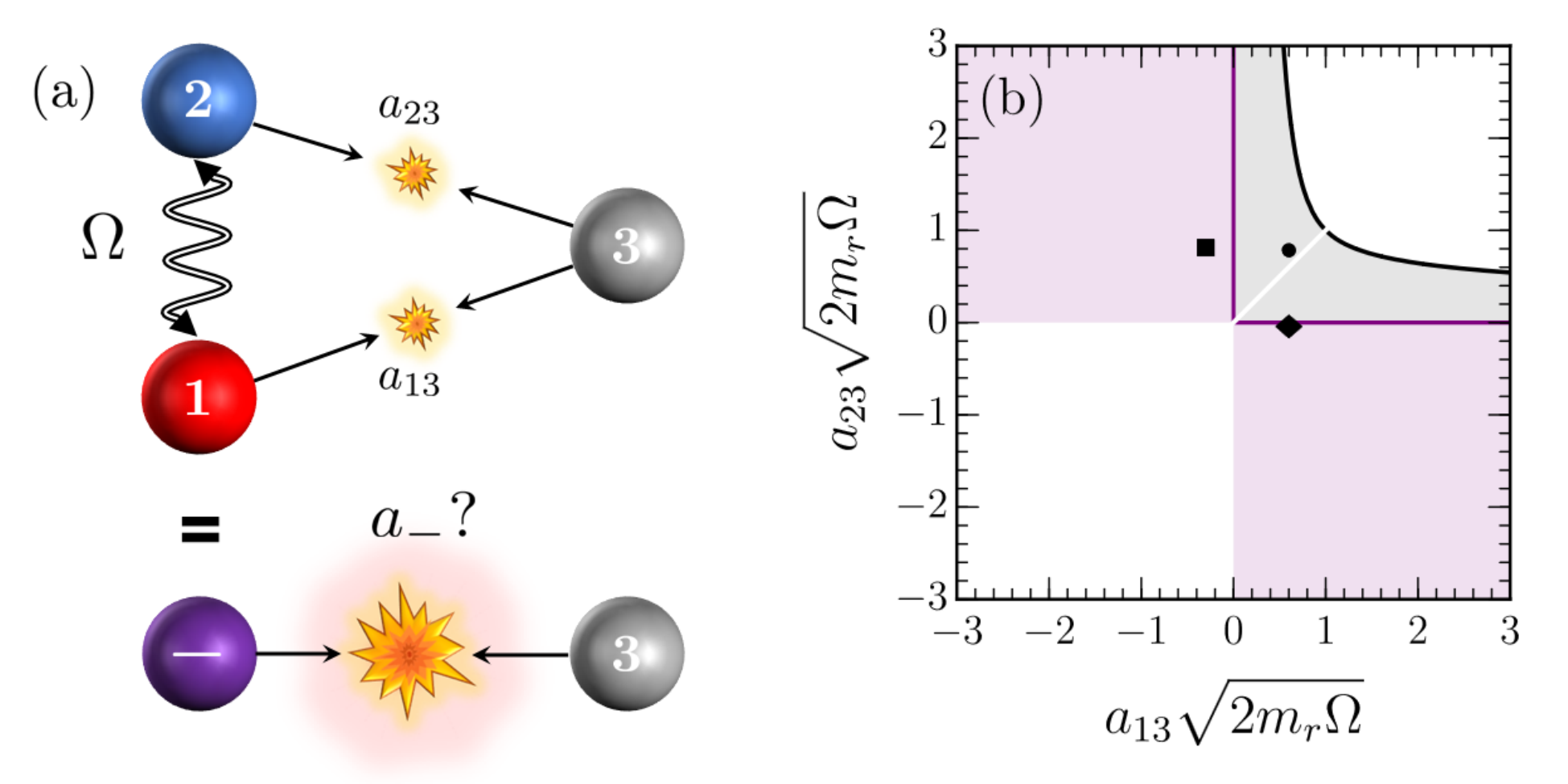}
\caption{(a) Sketch of the system considered. A coherent drive of strength $\Omega$ couples two atomic species (blue and red), which gives rise to dressed particles (purple).
Both of the Rabi coupled species can interact with a third component (gray) via short range interactions that are characterized by the scattering lengths $a_{13}$ and $a_{23}$. The resulting effective scattering length $a_-$ for the lowest energy dressed particle can exhibit resonances.
(b) Existence diagram of the dressed Feshbach resonances. 
The purple-shaded area indicates where a single resonance exists when varying the detuning of the Rabi drive, while in the gray area two resonances exist. By contrast, no resonances exist in the white areas, including along the diagonal where $a_{13}=a_{23}$. Here, $m_r$ is the reduced mass of the coupled and uncoupled components.}
\label{fig:sketch&exist}
\end{figure}

In this work, we combine these two concepts and investigate the effect of a strong Rabi drive on the interactions in a Fermi gas. Specifically, we consider a system where two hyperfine states of the same atom are coupled via a coherent drive, as illustrated in Fig.~\ref{fig:sketch&exist}(a), and the resulting dressed state is interacting with another hyperfine state or atomic species. This scenario thus requires three distinct states, although a strong coupling can effectively reduce it to a two-component system. Three-component Fermi gases have been experimentally realized with $^6$Li atoms~\cite{Ottenstein2008,Huckans2009,Williams2009,Lompe2010,schumacher2023}, and there are theoretical studies suggesting that radiofrequency (rf) radiation can be used to manipulate Feshbach resonances in this system~\cite{Zhang2009,Hanna_2010}. However, the previous calculations have all involved a multi-channel formulation, whereas here we use a minimal low-energy description which clearly exposes the key physics and allows us to extract analytic two-body parameters that can straightforwardly be applied to the many-body system.

Focusing on the scattering properties of the lowest-energy Rabi-dressed atom and an undriven atom [Fig.~\ref{fig:sketch&exist}(a)], we obtain an exact analytic expression for the effective scattering length characterizing this process, and we find that it can exhibit resonances for a wide range of parameters [Fig.~\ref{fig:sketch&exist}(b)]. Furthermore, we show that the resonances are linked to the emergence of hybrid two-body bound states, whose composition are strongly influenced by the drive. 
Our results thus demonstrate that the Rabi drive can significantly alter the two-body interactions, and in particular that resonances can be accessed purely by changing the parameters of the drive. A key advantage of the Rabi-drive-induced interactions over conventional magnetically tunable Feshbach resonances is that it gives access to an additional observable---the pseudospin of the coupled atoms---which can be used to monitor both few- and many-body physics. 

Exploiting the control of two-body interactions, we investigate the influence on the pairing instability---a precursor to the superfluid transition---in the Rabi-coupled Fermi gas. Within the Thouless criterion~\cite{Thouless1960,Strinati2018}, we find that one can achieve a crossover between different pairing regimes purely by varying the detuning of the Rabi drive, and we reveal an excited many-body branch which may be dynamically accessible in an out-of-equilibrium experiment.

\paragraph{Model and $T$ matrix.---}
We model the Rabi-coupled Fermi system depicted in Fig.~\ref{fig:sketch&exist}(a) with the Hamiltonian 
\begin{align} \label{eq:Ham} 
 \hat{H}&= \hat{H}_0+\hat{V}_1 + \hat{V}_2,
\end{align}
where
\begin{subequations}
\begin{align} 
\label{eq:Ham0} 
 \hat{H}_0&=\sum_{\k, j}\epsilon_{\k  j } \hat{f}_{\k j}^\dagger \hat{f}_{\k j}+ \frac{\Omega}{2} \sum_\k (\hat{f}_{\k 1}^\dagger \hat{f}_{\k 2}+h.c.),
 \\ \label{eq:V} 
  \hat{V}_j&= g_{j3} \sum_{\k,\k',\q} \hat{f}_{\k+\q j}^\dagger \hat{f}_{\k  j}\hat{f}_{\k'-\q 3}^\dagger  \hat{f}_{\k' 3} .
\end{align}
\end{subequations}
Here, $\hat{f}_{\k j}$ ($\hat{f}_{\k j}^{\dagger}$) are annihilation (creation) operators for fermions of species $j=\{1,2,3\}$ with momentum $\k$, and $\epsilon_{\k j }$ the corresponding kinetic energies. In the following, we take $\epsilon_{\k 1 }\equiv \epsilon_{\k  }= k^2/2m$, $\epsilon_{\k 2 } =\epsilon_\k+\delta$ and $\epsilon_{\k 3 }= k^2/2m_3$. Here, $\delta$ encodes the detuning of the Rabi drive frequency with respect to the 1-2 transition, and $\Omega$ is the strength of the coupling, where we have used the rotating wave approximation. We have also defined different masses $m$ and $m_3$ of the two coupled species and species 3, respectively, which allows us to investigate general mixtures. Finally, $g_{j3}$ corresponds to the ``bare'' strength of the contact interactions between species $j$ and 3. The bare couplings are related to the corresponding $s$-wave scattering lengths $a_{j3}$ via $\frac{1}{g_{j3}} = \frac{m_r}{2\pi a_{j3}} - \sum_{\k}^{\Lambda} \frac{1}{\epsilon_{\k}+\epsilon_{\k3}}$ with reduced mass $m_r=m m_3/(m+m_3)$, as well as an ultraviolet cutoff $\Lambda$ which will eventually be sent to infinity at the end of the calculation. Throughout, we work in units in which the volume, the reduced Planck constant, and Boltzmann's constant are all 1.

\begin{figure}    
\centering
\includegraphics[width=0.95\linewidth]{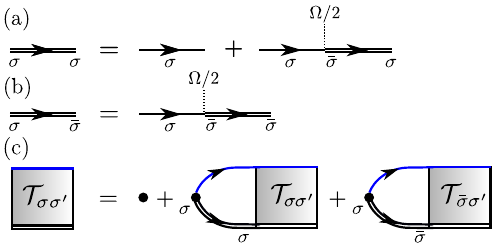}
    \caption{(a) Diagonal $G_{\sigma\sigma}$ and (b) off-diagonal $G_{\sigma\bar{\sigma}}$ $(\bar\sigma\neq\sigma)$ single-particle Green's functions for species $\sigma=\{1,2\}$ in the presence of a Rabi drive. The thin solid lines represent the bare Green's functions in the absence of Rabi drive, the vertical dotted lines illustrate the Rabi coupling between species 1 and 2, and the double lines represent the Rabi-dressed Green's functions. (c) Diagrammatic representation of the $T$-matrix equation, where the circle represents the interaction vertex $g_{\sigma 3}\delta_{\sigma\sigma'}$, and the thin blue line is the Green's function of a particle from species 3.
    }
    \label{fig:Greens_diagram}
\end{figure}

The noninteracting part of the Hamiltonian can be diagonalized as
\begin{align}
\hat{H}_0=\sum_{\k, \pm}(\epsilon_{\k } +\epsilon_{\pm})\hat{f}_{\k \pm}^\dagger \hat{f}_{\k \pm}+\sum_{\k }\epsilon_{\k 3 } \hat{f}_{\k 3}^\dagger \hat{f}_{\k 3}.    %
\end{align}
The dressed particle operators are related to the bare operators via
\begin{align}
\begin{pmatrix}
  \hat{f}_{\k-}\\
 \hat{f}_{ \k +}
 \end{pmatrix}&=\begin{pmatrix}
  c &&-s\\
s &&  c \end{pmatrix} \begin{pmatrix}
   \hat{f}_{\k 1}\\
   \hat{f}_{\k 2}
 \end{pmatrix},
\end{align}
where the transformation coefficients satisfy $c^2=\frac{1}{2}(1+\frac{\delta}{\sqrt{\delta^2+\Omega^2}})$, $cs=\frac{\Omega}{2\sqrt{\delta^2+\Omega^2}},$ and $c^2+s^2=1$. The dressed single-particle kinetic energies are now shifted parabolas $\epsilon_{\k } +\epsilon_{\pm}$, with $\epsilon_{\pm} =\frac{1}{2}\left(\delta\pm \sqrt{\delta^2+\Omega^2}\right)$.
Since the single-particle states of species 1 and 2 are no longer eigenstates of the non-interacting Hamiltionian, the corresponding single-particle Green's functions 
\begin{align}
G^{}_{\sigma\sigma'}(\k,\omega)=\bra{0}\hat{f}_{\k\sigma'} (\omega -\hat{H}_0+i0)^{-1}\hat{f}_{\k\sigma}^\dagger\ket{0},
\end{align}
with $\sigma,\sigma'=\{1,2\}$, are not diagonal and depend on the Rabi drive (the factor $+i0$ shifts poles infinitesimally into the lower half place). The diagrams corresponding to the Rabi-dressed Green's functions are presented in Fig.~\ref{fig:Greens_diagram}(a,b). By contrast, the third species is unaffected by the Rabi drive and thus its single-particle Green's function is simply given by $G_{3}(\k,\omega)%
=(\omega-\epsilon_{\k3}+i0)^{-1}$. 

We now turn to the interactions between an atom of species 3 with a Rabi-coupled atom, Fig.~\ref{fig:sketch&exist}(a).
The Rabi-induced interconversion processes discussed above must be accounted for, since it can couple bound and scattering states. 
As a result, we find that the two-body $T$-matrix that describes the interactions %
can be conveniently written as a $2\times2$ matrix equation of the form
\begin{align} \label{eq:tmat2by2}
 \mathbf{T}(\omega) =\mathbf{g}+\mathbf{g}\mathbf{\Pi}(\omega)   \mathbf{T}(\omega) ,
\end{align}
as illustrated in Fig.~\ref{fig:Greens_diagram}(c). In the basis of the pseudospins of species 1 and 2, we have 
\begin{align} 
\mathbf{g}&=\begin{pmatrix}
    g_{13} &0 \\ 0 &g_{23}
\end{pmatrix},~~
\mathbf{\Pi}(\omega) =\begin{pmatrix}
    \Pi_{11}(\omega) &\Pi_{12}(\omega)  \\ \Pi_{21}(\omega)  &\Pi_{22}(\omega)  
\end{pmatrix},
\end{align}
and the functions $\Pi_{\sigma\sigma'}(\omega)$ are related to the single-particle Green's functions as
\begin{align}
 \Pi_{\sigma\sigma'}(\omega)&= \new{i}\int \frac{d\omega'}{2 \pi}\sum_\q G^{}_{\sigma\sigma'}(-\q,-\omega') G_{3}(\q,\omega+\omega').
\end{align}
Inverting \eqref{eq:tmat2by2} gives $\mathbf{T}(\omega) =\left[\mathbf{g}^{-1}-\mathbf{\Pi}(\omega)\right]^{-1} $,
and the elements of the matrix can be evaluated analytically --- for details, see the Supplemental Material (SM)~\cite{supmatshort}. 

\new{Our model Hamiltonian \eqref{eq:Ham} does not include interactions between species 1 and 2. This is motivated by the fact that these are irrelevant for the two-body physics we discuss here and below, since it involves a single atom in a superposition of the Rabi-driven 1-2 components.
Furthermore, we expect that a small $a_{12}$ will not qualitatively affect our results regarding the pairing instabilities at finite density as long as the system remains in the strongly driven regime $\Omega \gtrsim E_F$, where the pairing between species 1 and 2 can be neglected.}

\begin{figure}[tbp] 
    \includegraphics[width=0.9\linewidth]{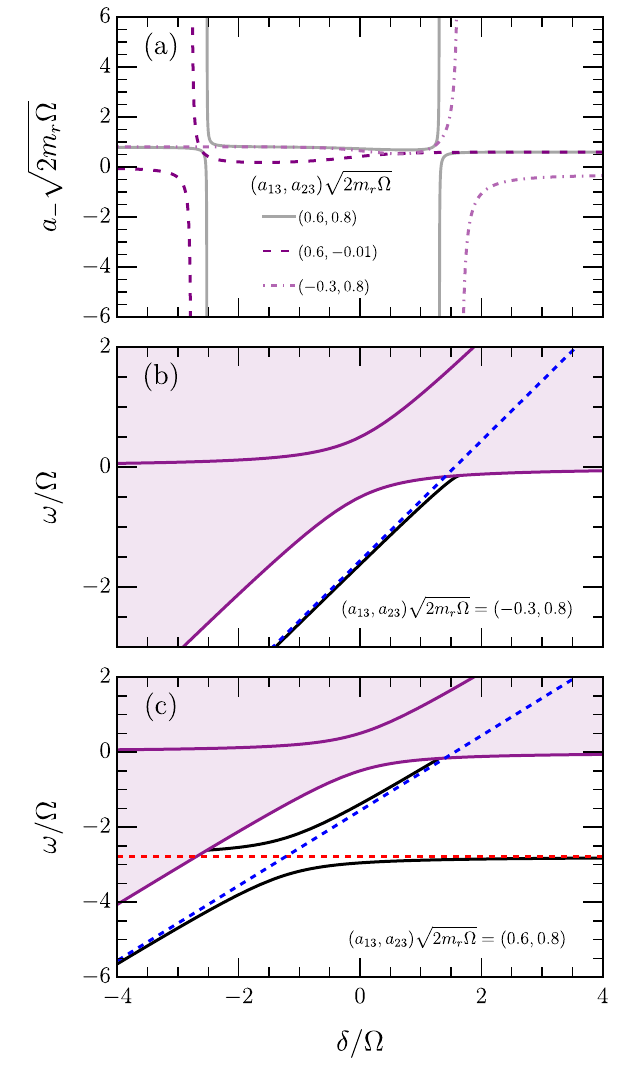}
\caption{Rabi-dressed Feshbach resonances and hybrid bound states as a function of detuning $\delta/\Omega$. (a) Dressed scattering length for three sets of bare scattering lengths ($a_{13},a_{23}$) corresponding to the points marked by ($\bullet,\blacklozenge,\blacksquare$) symbols in Fig.~\ref{fig:sketch&exist}(b). (b,c) Energy spectrum for $(a_{13},a_{23})\sqrt{2m_r\Omega}=(-0.3,0.8)$ in (b), and $(a_{13},a_{23})\sqrt{2m_r\Omega}=(0.6,0.8)$ in (c). We show the hybrid bound states (black lines), with the dashed red and blue lines corresponding to the bare 1-3 and 2-3 bound states. The purple lines show the dressed single-particle energies $\epsilon_{\pm}$ and the shaded area highlights the scattering continuum of the driven system.}
\label{fig:DressScatt}
\end{figure}

\paragraph{Rabi-drive-induced Feshbach resonances.---}
We now demonstrate how the application of a strong Rabi drive can lead to scattering resonances that can be accessed purely by varying the parameters of the drive, such as the detuning, rather than by varying a magnetic field as in conventional Feshbach resonances. To this end, we consider the elastic scattering between a lower dressed particle $\hat{f}^\dagger_{\k-}$ and a particle in the third component (as we explain below, scattering involving the upper dressed state does not lead to a resonance). Naively, one might expect that the effective scattering length for this process is given by a weighted sum of the bare scattering lengths of the form $a_{13}c^2+a_{23}s^2$, which only features a resonance if one of the bare scattering lengths is resonant. Indeed, this approximation works well when one considers weakly interacting Bose gases~\cite{Search2001,Lavoine2021,Sanz2022}. However, it neglects the interconversion between species 1 and 2 during the scattering process, as well as the shift of the single-particle continuum. The possibility of resonances depends on the non-trivial interplay between these effects.

To obtain the scattering $T$ matrix between the lower dressed state and an atom of species 3, we rotate into the dressed-state basis which gives $T_{-}(\omega)= c^2 \mathbf{T}_{11}(\omega)+s^2\mathbf{T}_{22}(\omega)-2cs\mathbf{T}_{12}(\omega)$, where we have used $\mathbf{T}_{21}(\omega)=\mathbf{T}_{12}(\omega)$. In the limit of small collision energy $\epsilon_\k+\epsilon_{\k3}$, measured from the shifted continuum $\epsilon_{-}$, this can be expanded as 
\begin{align}\label{eq:LowkTdressed}
T_{-}^{-1}(\epsilon_{-}+\epsilon_\k+\epsilon_{\k3}) \simeq \frac{m_r }{2\pi}\left(\frac{1}{a_{-}} - r_{\text{eff}}\frac{k^2}{2} +i k \right),
\end{align}
where we have introduced the dressed scattering length $a_{-}$ and the effective range $r_{\text{eff}}$. We find that these can be expressed as
\begin{subequations}
\begin{align}
 \label{eq:scattlength1323}
\frac{1}{a_{-}}&= \frac{1-\sqrt{2 m_r(\epsilon_{+}-\epsilon_{-})} (a_{13}s^2+a_{23}c^2)}{a_{13}c^2+a_{23}s^2-a_{13}a_{23}\sqrt{2 m_r(\epsilon_{+}-\epsilon_{-})}},
\\  \label{eq:efferange1323}
r_{\text{eff}}&=\frac{-c^2s^2(a_{13}-a_{23})^2
/\sqrt{2 m_r(\epsilon_{+}-\epsilon_{-})} }{\left[a_{13}c^2+a_{23}s^2-a_{13}a_{23}\sqrt{2 m_r(\epsilon_{+}-\epsilon_{-})} \right]^2}.
\end{align}
\end{subequations}
In particular, we see that in general $a_{-}\neq a_{13}c^2+a_{23}s^2$, and that the effective range $r_{\text{eff}}\leq0$, similar to the case for a standard two-channel model of a magnetically tunable Feshbach resonance~\cite{Timmermanns1999,Gurarie2007}.
We also note that when $a_{23}=0$, Eq.~\eqref{eq:scattlength1323} reduces to the effective scattering length derived in Refs.~\cite{mulkerin2024,Zulli2025}. 
In the high-symmetry scenario where $a_{13}=a_{23}=a$, Eq.~\eqref{eq:scattlength1323} reduces to $a_{-}=a$\new{. This can be understood from the fact that, in this limit, the parts of the Hamiltonian corresponding to the lower and upper dressed fermions decouple \cite{supmatshort}. T}hus for this special case, the dressed scattering length is independent of the Rabi drive and only diverges when $a$ is itself divergent~\footnote{Indeed, when $a_{13}=a_{23}=a$, the $T$ matrix reduces to that in the absence of Rabi coupling, independent of the parameters of the Rabi drive, with $T_{-}(\epsilon_-+\epsilon_\k+\epsilon_{\k 3})=\frac{2\pi}{m_r}\frac1{a^{-1}+ik}$}.

When $a_{13}\neq a_{23}$, however, $a_{-}$ is strongly affected by the drive and can exhibit new resonances when the numerator of Eq.~\eqref{eq:scattlength1323} vanishes. Depending on the values of $a_{13}$ and $a_{23}$, we find that this can occur for either zero, one, or two values of the dimensionless parameter $\delta/\Omega$. Figure~\ref{fig:sketch&exist}(b) shows the existence diagram of the Rabi-induced resonances. The shaded areas are regions where resonances exist, while white denotes regions without a resonance. In the purple regions there exists a single resonance, whereas the gray area contains two resonances. To connect these regions, we note that on the horizontal (vertical) purple line boundary, a single resonance exists at negative (positive) detuning. When moving infinitesimally from the horizontal (vertical) axis into the gray area, a second resonance enters from  $\delta\rightarrow \pm \infty$ and as we move further away from the axis %
along a vertical (horizontal) line, the second resonance approaches the initial resonance from above (below), and the two resonances eventually merge and vanish when $\Omega$ is sufficiently large 
(denoted by the black line).

Figure \ref{fig:DressScatt}(a) shows the dressed scattering length and associated resonances, calculated for three sets of bare scattering lengths %
that are located in each colored area of the existence diagram, as indicated by the symbols ($\bullet,\blacklozenge,\blacksquare$) in Fig.~\ref{fig:sketch&exist}(b). We see that the dressed scattering length exhibits a single resonance when only one of the bare scattering lengths is positive, while there are two resonances when
both of the bare scattering lengths are positive, 
as expected from the existence diagram and the discussion above. 
One can also understand the behavior of $a_{-}$ in the vicinity of the resonances, which we find can be approximated as
\begin{subequations}
   \begin{align}\label{eq:approxdressedscat}
   a_{-}&\simeq  a_{\text{bg} }\left(1+\frac{\Gamma}{\delta-\delta_c}\right),\\
    \Gamma&=  \frac{1}{a_{\text{bg} }} \frac{  c^2 s^2  (\epsilon_{+}-\epsilon_{-}) (a_{13}-a_{23})^2  }{\frac{c^2-s^2}{2\sqrt{2m_r(\epsilon_{+}-\epsilon_{-})}}+a_{13}s^2-a_{23}c^2} ,
   \end{align}
\end{subequations}
where we have introduced an effective background scattering length $a_{\text{bg}}=a_{13}c^2+a_{23}s^2$, and $\epsilon_{\pm},c,s$ are evaluated at the critical detuning $\delta_c$. Hence, we expect the resonance width $\new{|}\Gamma\new{|}$ to be reduced when $|a_{13}-a_{23}|$ decreases, as can be observed in Fig.~\ref{fig:DressScatt}(a), and to be enhanced when $\Omega$ increases. \new{We note that $\Gamma$ can be either positive or negative, and that both cases appear in Fig.~\ref{fig:DressScatt}(a) where we can observe inverted resonance shapes on the negative and positive detuning sides.} Interestingly, we see that Eq.~\eqref{eq:approxdressedscat} resembles the typical two-channel behavior of a Feshbach resonance \cite{Moerdijk1995,Inouye1998,Chin2010}, although in the Rabi-driven case,  $a_{\text{bg}}$ also depends on the detuning.
It also has the same form as the real part of the (inverse) effective scattering length in the presence of an oscillating magnetic field \cite{Smith2015}. However, in our scenario the scattering length is purely real, and unlike the scenario in Ref.~\cite{Smith2015} the background scattering length itself depends on the applied field.

The scattering of the upper dressed state and an atom of species 3 is obtained in a similar manner, with $a_{+}=(m_r/2\pi)\left[s^2 \mathbf{T}_{11}(\epsilon_+)+c^2\mathbf{T}_{22}(\epsilon_+)+2cs\mathbf{T}_{12}(\epsilon_+)\right]$.
We obtain
\begin{align}\label{eq:a+}
\frac{1}{a_{+}}&= \frac{1+i\sqrt{2 m_r(\epsilon_{+}-\epsilon_{-})} (a_{13}c^2+a_{23}s^2)}{a_{13}s^2+a_{23}c^2+i a_{13}a_{23}\sqrt{2 m_r(\epsilon_{+}-\epsilon_{-})}}.
\end{align}
In contrast to $a_{-}$, it does not exhibit any resonance and it is in general complex valued since the scattering takes place in the dressed continuum (except if $a_{13}=a_{23}=a$, in which case it reduces to $a_{+}=a$). %
The effect of the continuum is less relevant when $\sqrt{2 m_r(\epsilon_{+}-\epsilon_{-})} (a_{13}c^2+a_{23}s^2)\ll1$ where we have $a_{+}\simeq (a_{13}s^2+a_{23}c^2)$\new{, or in the limits $\delta/\Omega\rightarrow\pm \infty$ where $a_{+}$ reduces to $a_{23}$ or $a_{13}$}.

While we focus on fermions in this work, our few-body results apply to atoms of any statistics. Previous works have investigated the Feshbach resonances of rf-dressed bosonic atoms~\cite{Moerdijk1996,Kaufman2009,Tscherbul2010,Papoular2010,Hanna_2010} (see also Ref.~\cite{Petrov2014}), but these typically involved scattering between identical bosons, unlike the scenario we consider where only one of the colliding particles is dressed. Furthermore, many proposals involve the coupling to a deeply bound molecular state, similar to the situation in an optical Feshbach resonance~\cite{Fedichev1996}. 
By contrast, our work uses a low-energy model featuring shallow bound states that can be strongly modified by the Rabi drive.

\paragraph{Hybrid bound states.---}
The origin of the dressed resonances can be understood through the two-body bound states of the Rabi-coupled system.
The energy of these {\it hybrid} bound states can be calculated from the (real) poles of the $T$ matrix (see the SM~\cite{supmatshort} for an equivalent solution starting from the \sch equation).

Figure \ref{fig:DressScatt}(b) illustrates the case where only one scattering length is positive. In this case,  we find a single bound state below the continuum (shaded purple area starting at $\epsilon_-$). At negative detuning the bound state approaches the bare bound state energy, $-1/(2m_ra_{23}^2)+\delta$, from below. On the other hand, when the hybrid bound state approaches the continuum at positive detuning, its energy is modified.
The detuning at which the bound state enters the continuum corresponds to the position of a resonance in Fig.~\ref{fig:DressScatt}(a).

When $a_{13}$ and $a_{23}$ are both positive, we instead have two hybrid bound states, as shown in Fig.~\ref{fig:DressScatt}(c). In particular, we find that the upper and lower solutions involve 1-3 and 2-3 dimers that are either in or out of phase~\cite{supmatshort}. Away from the continuum, the bound-state energies interpolate between the bare bound states with a clear anticrossing, and they are well approximated by 
\begin{align}\label{eq:boundstatehybrid}
  \omega_{\pm}\simeq\frac{1}{2}\left(-\bar{\varepsilon}+\delta  \pm \sqrt{(\varepsilon_{23}-\varepsilon_{13}-\delta)^2+\Omega^2} \right),
\end{align}
with $\bar{\varepsilon}=\varepsilon_{13}+\varepsilon_{23}$ and $\varepsilon_{ij}=1/(2m_r a_{ij}^2)$. Here, $\omega_{-}$ corresponds to the ground state, which always exists, while $\omega_{+}$ approximates the excited bound state and is only a true bound state outside of the continuum. 
Indeed, as we can see in Fig.~\ref{fig:DressScatt}(c), the excited bound state disappears when entering the continuum, and the two values of detunings at which this occurs correspond to the positions of resonances of $a_{-}$  in Fig.~\ref{fig:DressScatt}(a).
In the special case $a_{23}=a_{13}=a$, we find that Eq.~\eqref{eq:boundstatehybrid} becomes exact with $ \omega_{\pm}=\epsilon_{\pm}-1/(2m_r a^2)$. Here, the excited hybrid bound state remains decoupled from the continuum of lower dressed particles, which explains the absence of a dressed Feshbach resonance mentioned above.

\begin{figure}[tbp] 
    \includegraphics[width=0.9\linewidth]{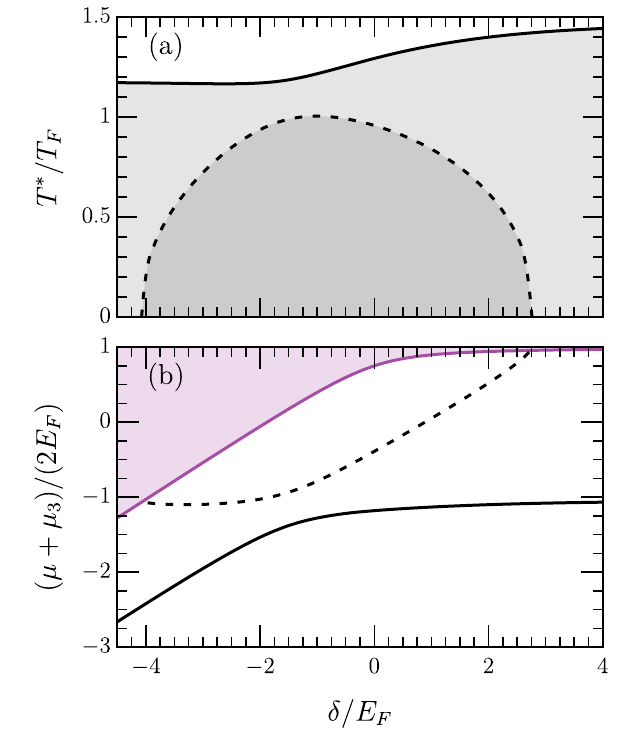}
\caption{(a) Critical temperatures $T^*$ for pairing at equal densities $n_1 + n_2 = n_3$, and (b) associated average chemical potentials versus detuning for $(a_{13},a_{23})\sqrt{2m_r\Omega}=(0.6,0.8)$, $\Omega/E_F=1$ and equal masses $m_3=m$. The black solid (dashed) lines correspond to 
ground (excited) state solutions.
The purple line in (b) marks the edge of the dressed many-body continuum, given by $2E_F+\epsilon_{-}$.}
\label{fig:Tc}
\end{figure}

\paragraph{Pairing instabilities in a Rabi-coupled Fermi gas.---}
The results presented above have implications beyond two-body physics. Indeed, considering Fermi gases with a finite density, it is clear that the interactions allow for pairing to take place, which can give rise to a transition to a superfluid state. In conventional uncoupled Fermi systems, it is known that the pairing instability at finite temperature can be related to the many-body $T$ matrix through the so-called Thouless criterion \cite{Thouless1960,Strinati2018}. We obtain a generalization of this criterion \cite{Longpaper} for the Rabi-coupled system, which takes the form 
\begin{align}\label{eq:ThoulessCrit}
 \det [\mathbf{T}^{\text{mb}}(0)^{-1}]=0,
\end{align}
where $\mathbf{T}^{\text{mb}}(\omega)$ denotes the many-body $T $ matrix. It consists of the same diagrams as the two-body case [see Fig.~\ref{fig:Greens_diagram}(c)], adjusted to include the effect of Pauli exclusion due to the finite fermion density. Thus, $\mathbf{T}^{\text{mb}}$ is given by Eq.~\eqref{eq:tmat2by2} with $\mathbf{\Pi}$ replaced by its many-body counterpart, which accounts for this effect through the Fermi occupations of the dressed fermions $n_{\q\pm}=1/(e^{\beta (\epsilon_{\q}+\epsilon_{\pm}-\mu)}+1)$ and the third component $n_{\q3}=1/(e^{\beta (\epsilon_{\q3}-\mu_3)}+1)$, with the inverse temperature $\beta\equiv1/T$ and chemical potentials $\mu$ and $\mu_3$ describing the densities $n_1+n_2$ and $n_3$, respectively \cite{supmatshort}.

Solving Eq.~\eqref{eq:ThoulessCrit} and the number equations self consistently allows us to determine the critical temperature for pairing (which is directly connected to the superfluid transition in the weak-coupling BCS regime) and the chemical potentials. 
Interestingly, we find that, similarly to the hybrid bound states in the two-body problem, there are regimes where two distinct pairing instabilities can occur, corresponding to 1-3 and 2-3 pairs that are either in phase or out of phase. The results of such a calculation for equal masses $m=m_3$ and balanced densities $n_{1}+n_{2}=n_{3}\equiv k_F^3/6\pi^2$ are presented in Fig.~\ref{fig:Tc}, where we define the Fermi energy $E_{F}=k_{F}^2/2m$ and corresponding Fermi temperature $T_F=E_F$.
We observe in Fig.~\ref{fig:Tc}(a) that a critical temperature always exists for the lowest energy many-body paired state (solid black line), and likewise there is a well-defined average chemical potential in Fig.~\ref{fig:Tc}(b).
This suggests that the corresponding zero temperature ground state exhibits a superfluid \textit{crossover} where the pairing evolves continuously from being dominated by the 2-3 channel at negative detuning to being dominated by 1-3 pairing at positive detuning. For the parameters chosen here where $a_{13},a_{23}>0$, this can be viewed as a BEC-BEC crossover, where the pseudospin of the pairs smoothly changes sign.

Additionally, Fig.~\ref{fig:Tc}(a) shows that there also exists an excited many-body paired state that vanishes at some critical detunings. %
This behavior is reflected in panel (b), which shows how the corresponding average chemical potentials
reaches the Rabi-dressed many-body continuum %
at the same critical detunings.
This signals that the excited state can undergo a transition from a superfluid to normal phase at these points. The excited-state superfluid can possibly be accessed dynamically by manipulating the bare scattering lengths such that the lower bound state is initially not present, i.e., by starting in the purple region in the vicinity of one of the purple lines in Fig.~\ref{fig:sketch&exist}(b) and tuning across the purple line into the gray region with two resonances. As a function of detuning, this excited state experiences normal state-BCS transitions at the critical detunings where the chemical potential becomes that of two noninteracting Fermi seas, as well as a BCS-BEC-BCS crossover in between~\cite{Longpaper}.

We stress that we have calculated the critical temperature using the number equations of fermions in the normal state, and thus, strictly speaking, it corresponds to the critical temperature for fermion pairing to take place. While pairing is essential for the emergence of fermionic superfluids, it is well known that this criterion overestimates the transition temperature in the regime where the pairs are tightly bound, since it exceeds the condensation temperature of the ideal Bose gas of dimers \cite{Strinati2018,Nozieres1985,SadeMelo1993}. This should be kept in mind for the interpretation of the large values of $T^*$ obtained in Fig.~\ref{fig:Tc}. Nonetheless, our results clearly illustrate that one can tune across the entire BCS-BEC crossover by varying the parameters of the Rabi drive alone.

Finally, we comment on the stability of the Rabi-coupled Fermi gas. In general, strongly interacting three-component Fermi gases are unstable towards three-body recombination~\cite{Ottenstein2008,Lompe2010}. While this is not an issue for the two-body physics described above, as it only involves two particles, the Rabi-dressed many-body system necessarily features all three bare fermionic components. However, we note that, in our setup, the \new{negligible} 1-2 interactions strongly suppresses both Efimov physics and three-body losses compared with the case where all components interact near resonantly~\cite{Braaten2006}. Additional stability can be achieved by considering $\Omega\gtrsim E_F,T$ such that the population of the upper dressed state remains small and we are left with an effective two-component Fermi gas. Indeed, the polaron limit of the strongly Rabi-driven Fermi gas was investigated in Ref.~\cite{Vivanco2025}, and despite the significant impurity fraction of 1:5 the authors did not find that losses noticeably impacted their measurements.

\paragraph{Conclusions.---}
We have shown that the presence of a Rabi drive can strongly modify the interactions in ultracold atomic gases. Using a $T$-matrix calculation, we have demonstrated that the scattering between a Rabi-driven and an undriven atom can exhibit resonances, which arise when hybrid bound states enter the Rabi-dressed scattering continuum. Beyond the two-body physics, we find that these results have implications for the pairing instabilities in the Rabi-coupled Fermi gas. Specifically, we have shown that for a given set of bare scattering lengths, one can use the detuning to control the transition temperature and tune the system across superfluid crossovers. We expect that our findings will be relevant for other few and many body phenomena in Rabi-coupled mixtures, such as Efimov states \cite{Zulli2025} or ground-state transitions in spin-imbalanced gases~\cite{Mora2009,Punk2009,Combescot2009}. The investigation of these mixtures may also shed light on other physical systems difficult to access experimentally. For example, the three-component Fermi gas is known to have connections with quark matter \cite{Rapp2007,Wilczek2007,Ozawa2010,O'Hara_2011}, and similar analogies may exist in the Rabi-driven case \cite{Kurkcuoglu2018}. Furthermore, it could provide a useful analog of light-driven materials, allowing one to simulate phenomena such as light-enhanced superconductivity~\cite{Fausti2011,Mitrano2016}.

\begin{acknowledgments}
We gratefully acknowledge fruitful discussions with Henning Moritz, Nir Navon, and Anthony Zulli.
OB, BCM, JL, and MMP acknowledge support from the Australian Research Council (ARC) Centre of Excellence in Future Low-Energy Electronics Technologies (CE170100039), and from ARC Discovery Projects DP240100569 and DP250103746. 
MMP is also supported through an ARC Future Fellowship FT200100619. CRC is supported by the Cluster of Excellence “CUI: Advanced Imaging of Matter”—EXC 2056—project ID 390715994.
\end{acknowledgments}

{\it Data Availability.}---
The data that support the findings of this article are openly available \cite{short_data}.

\bibliography{biblio}

\renewcommand{\theequation}{S\arabic{equation}}
\renewcommand{\thefigure}{S\arabic{figure}}

\onecolumngrid

\newpage

\setcounter{equation}{0}
\setcounter{figure}{0}
\setcounter{page}{1}

\clearpage
\appendix
\section{SUPPLEMENTAL MATERIAL: Scattering resonances and pairing in a Rabi-coupled Fermi gas}

\begin{center}
Olivier Bleu,$^1$
Brendan C. Mulkerin,$^1$
Cesar R. Cabrera,$^{2,3}$
Jesper Levinsen,$^1$
Meera M. Parish$^1$\\
\emph{\small $^1$School of Physics and Astronomy, Monash University, Victoria 3800, Australia}\\
\emph{\small $^2$Institute for Quantum Physics, Universität Hamburg, Luruper Chaussee 149, 22761 Hamburg, Germany\\
$^3$The Hamburg Centre for Ultrafast Imaging, Universität Hamburg, Luruper Chaussee 149, 22761 Hamburg, Germany}
\end{center}

\new{\subsection{Absence of Rabi-induced resonances when $a_{13}=a_{23}$}

In this section, we show that the absence of resonances when the bare scattering lengths are equal can be understood from the symmetry of the Hamiltonian.
To visualize this, it is convenient to write the interaction part of the Hamiltonian in terms of the dressed fermions operators:
\begin{subequations}
\begin{align}\nonumber
    \hat{V}&=\hat{V}_1+\hat{V}_2 \\
    &= g_{13}\sum_{\k,\k',\q} \hat{f}_{\k+\q 1}^\dagger \hat{f}_{\k  1}\hat{f}_{\k'-\q 3}^\dagger  \hat{f}_{\k' 3}+g_{23}\sum_{\k,\k',\q} \hat{f}_{\k+\q 2}^\dagger \hat{f}_{\k  2}\hat{f}_{\k'-\q 3}^\dagger  \hat{f}_{\k' 3}
    \\ \nonumber
     &= (g_{13}c^2+g_{23}s^2) \sum_{\k,\k',\q} \hat{f}_{\k+\q -}^\dagger \hat{f}_{\k  -}\hat{f}_{\k'-\q 3}^\dagger  \hat{f}_{\k' 3}+(g_{13}s^2+g_{23}c^2) \sum_{\k,\k',\q} \hat{f}_{\k+\q +}^\dagger \hat{f}_{\k  +}\hat{f}_{\k'-\q 3}^\dagger  \hat{f}_{\k' 3} \\ \label{eq:interdressedbasis}
     & ~~~~~+cs (g_{13}-g_{23}) \sum_{\k,\k',\q} \left( \hat{f}_{\k+\q +}^\dagger \hat{f}_{\k  -}+\hat{f}_{\k+\q -}^\dagger \hat{f}_{\k  +}\right)\hat{f}_{\k'-\q 3}^\dagger  \hat{f}_{\k' 3},
\end{align}
\end{subequations}
where we have used the transformation given in Eq.~(4) of the main text.
As we can see from the second line in Eq.~\eqref{eq:interdressedbasis}, in the dressed basis, the interactions with the third species give rise to an effective coupling between the dressed fermions $\hat f_\pm$.
However, this coupling vanishes when $g_{13}=g_{23}=g$, and thus, in this special case, the three-component system can be effectively reduced to two two-component systems $\{-,3\}$ and $\{+,3\}$ with interaction strengths $g$.
Since the $\{-,3\}$ and $\{+,3\}$ scattering channels are now completely decoupled, the presence of a bound state in the $\{+,3\}$ channel cannot affect the scattering in the $\{-,3\}$ channel, and thus, it cannot give rise to new resonances. 
This is also why the dressed $T$-matrices reduce to $T_{\pm}(\omega)=\frac{2 \pi/m_r}{a^{-1}-\sqrt{\epsilon_{\pm}-\omega-i 0}}$, i.e., they take the same form as a standard bare $T$-matrix with a shifted continuum.}

\subsection{Details on the calculation of the $T$ matrix}
In this section, we provide additional details on the calculation of the $T$ matrix. As shown in the main text, the two-body $T$ matrix is of the form
\begin{align} \label{eq:tmat2by2SM}
   \mathbf{T}(\omega)& =\left[\mathbf{g}^{-1}-\mathbf{\Pi}(\omega)\right]^{-1},
\end{align}
with
\begin{align}
 \Pi_{\sigma\sigma'}(\omega)&= \new{i}\int \frac{d\omega'}{2 \pi}\sum_\q G^{}_{\sigma\sigma'}(-\q,-\omega') G_{3}(\q,\omega+\omega').
\end{align}
The non-interacting Green's function of the third species is given by $G_{3}(\k,\omega)=(\omega -\epsilon_{\k3}+i0)^{-1}$, and those of the Rabi-coupled species $G^{}_{\sigma\sigma'}(\k,\omega)=\bra{0}\hat{f}_{\k\sigma'} (\omega -\hat{H}_0+i0)^{-1}\hat{f}_{\k\sigma}^\dagger\ket{0}$ read~\cite{Hu2023Rabi,mulkerin2024}
\begin{subequations}
    \begin{align}
 G_{11}(\k,\omega)&=\frac{c^2}{\omega-\epsilon_{\k}-\epsilon_{-}}+ \frac{s^2}{\omega-\epsilon_{\k}-\epsilon_{+}},
\\
 G_{22}(\k,\omega)&=\frac{s^2}{\omega-\epsilon_{\k}-\epsilon_{-}}+ \frac{c^2}{\omega-\epsilon_{\k}-\epsilon_{+}} ,
\\
G_{12}(\k,\omega)&=G_{21}(\k,\omega) =\frac{-cs}{\omega-\epsilon_{\k}-\epsilon_{-}}+ \frac{cs}{\omega-\epsilon_{\k}-\epsilon_{+}}.
\end{align}
\end{subequations}
This allows us to evaluate the frequency integrals in the pair functions $\Pi_{\sigma\sigma'}$, which gives
\begin{subequations}\label{eq.Piij}
    \begin{align}
 \Pi_{11}(\omega)&=\sum_\q \left[\frac{c^2}{\omega-\bar{\epsilon}_{\q}-\epsilon_{-}}+ \frac{s^2}{\omega-\bar{\epsilon}_{\q}-\epsilon_{+}}\right],
\\
 \Pi_{22}(\omega)&=\sum_\q \left[\frac{s^2}{\omega-\bar{\epsilon}_{\q}-\epsilon_{-}}+ \frac{c^2}{\omega-\bar{\epsilon}_{\q}-\epsilon_{+}} \right],
\\
 \Pi_{12}(\omega)&= \Pi_{21}(\omega)=\sum_\q \left[\frac{-cs}{\omega-\bar{\epsilon}_{\q}-\epsilon_{-}}+ \frac{cs}{\omega-\bar{\epsilon}_{\q}-\epsilon_{+}}\right]
,
\end{align}
\end{subequations}
where we have introduced $\bar{\epsilon}_{\q}=\epsilon_{\q3}+\epsilon_\q$. The momentum integrals can then be evaluated analytically, and the elements of the inverse $T$ matrix read
\begin{subequations}
\begin{align} 
    \frac{1}{g_{13}}-\Pi_{11}(\omega) &=  \frac{m_r}{2\pi}\left[\frac{1}{a_{13}}-\left( c^2 \sqrt{2 m_r(\epsilon_{-}-\omega)}+s^2 \sqrt{2 m_r(\epsilon_{+}-\omega)}\right)\right], 
   \\ 
      \frac{1}{g_{23}}-\Pi_{22}(\omega)  &= \frac{m_r}{2\pi}\left[\frac{1}{a_{23}}-\left( s^2 \sqrt{2 m_r(\epsilon_{-}-\omega)}+c^2 \sqrt{2 m_r(\epsilon_{+}-\omega)}\right)\right],
 \\ 
 \Pi_{12}(\omega)   & =    \Pi_{21}(\omega) =cs \frac{m_r}{2\pi} \left[-\sqrt{2 m_r(\epsilon_{-}-\omega)}+\sqrt{2 m_r(\epsilon_{+}-\omega)}\right]. 
\end{align}
\end{subequations}

\subsubsection{Many-body $T$ matrix}

 The many-body $T$ matrix of the Rabi coupled Fermi gas has the same form as the two-body $T$ matrix but with $\mathbf{\Pi}$ replaced by its many-body version $\mathbf{\Pi}^{\text{mb}}$ which accounts for the Pauli blocking of identical fermions. 
\begin{align} 
   \mathbf{T}^{\text{mb}}(\omega)& =\left[\mathbf{g}^{-1}-\mathbf{\Pi}^\text{mb}(\omega)\right]^{-1} .
\end{align}
Specifically, the elements of $\mathbf{\Pi}^{\text{mb}}$ are  given by
\begin{subequations}
    \begin{align}
 \Pi_{11}^{\text{mb}}(\omega)&=\sum_\q \left[\frac{c^2(1-n_{\q3}-n_{\q-})}{\omega-\bar{\xi}_{\q}-\epsilon_{-}}+ \frac{s^2(1-n_{\q3}-n_{\q+})}{\omega-\bar{\xi}_{\q}-\epsilon_{+}}\right],
\\
 \Pi_{22}^{\text{mb}}(\omega)&=\sum_\q \left[\frac{s^2(1-n_{\q3}-n_{\q-})}{\omega-\bar{\xi}_{\q}-\epsilon_{-}}+ \frac{c^2(1-n_{\q3}-n_{\q+})}{\omega-\bar{\xi}_{\q}-\epsilon_{+}}\right] ,
\\
 \Pi_{12}^{\text{mb}}(\omega)&= \Pi_{21}^{\text{mb}}(\omega)=\sum_\q \left[\frac{-cs(1-n_{\q3}-n_{\q-})}{\omega-\bar{\xi}_{\q}-\epsilon_{-}}+ \frac{cs(1-n_{\q3}-n_{\q+})}{\omega-\bar{\xi}_{\q}-\epsilon_{+}}\right]
 .
\end{align}
\end{subequations}
Here, we have introduced the Fermi occupations of the Rabi-dressed fermions $n_{\q\pm}=1/(e^{\beta (\epsilon_{\q}+\epsilon_{\pm}-\mu)}+1)$ and of the third component $n_{\q3}=1/(e^{\beta (\epsilon_{\q3}-\mu_3)}+1)$ with the  chemical potentials $\mu$ and $\mu_3$. We have also defined $\bar{\xi}_{\q}=\epsilon_{\q}-\mu+\epsilon_{\q 3}-\mu_3$.

The associated number equations for the densities $n_1$, $n_2$ and $n_3$ of the three components in the many-body system are:
\begin{align}
    n_1 + n_2 = & \sum_\q \left( \frac{1}{e^{\beta (\epsilon_{\q}+\epsilon_{+}-\mu)}+1} + \frac{1}{e^{\beta (\epsilon_{\q}+\epsilon_{-}-\mu)}+1} \right), \\
    n_3 = & \sum_\q \frac{1}{e^{\beta (\epsilon_{\q3}-\mu_3)}+1} \, .
\end{align}
We use these number equations to calculate the pairing instability temperatures and the chemical potentials at a fixed density as shown in Fig.~4 of the main text.

\subsection{Bound states using the Schr\"{o}dinger equation}
Alternatively, we can obtain the two-body bound states by solving the Schr\"{o}dinger equation. This allows us to show that the bound-state energy can be obtained from the poles of the dressed $T$ matrix, as done in the main text, and, in addition, to derive the corresponding wave functions.

We are interested in calculating the two-body bound states between a dressed particle (superposition of  species 1 and 2) and a particle from species 3. 
The most general two-body wave function takes the form
\begin{align}\label{eq:2bodywavefunction}
 \ket{\psi}&=\sum_\k \varphi_\k \hat{f}_{\k3}^\dagger \left(c_\k \hat{f}_{-\k1}^\dagger+ s_\k \hat{f}_{-\k2}^\dagger\right)\ket{0}.
\end{align}
Note that $c_\k$ and $s_\k$ are part of the bound state wave function that we wish to determine and differ from the single-particle mixing coefficients. Projecting the two-body Schr\"{o}dinger equation $\hat{H} \ket{\psi}=E\ket{\psi}$ onto its components, we obtain
 \begin{subequations}  \label{eq:Schrodinger}
\begin{align} \label{eq:Schrodingera}
 \left(E-\bar{\epsilon}_\k\right) \varphi_\k c_\k -\frac{\Omega}{2}\varphi_\k s_\k &=  g_{13}\sum_\q \varphi_\q c_\q, 
\\   \label{eq:Schrodingerb} \left(E-\bar{\epsilon}_\k-\delta\right)\varphi_\k s_\k-\frac{\Omega}{2}\varphi_\k c_\k&=  g_{23}\sum_\q \varphi_\q s_\q.
\end{align}
 \end{subequations}
Taking the ratio of the two parts in Eq.~\eqref{eq:Schrodinger} we obtain an equation for $t_\k=s_\k/c_\k$
\begin{align} \label{eq:tk13232body}
t_\k = \frac{\Omega+2(E-\bar{\epsilon}_\k) R}{2(E-\bar{\epsilon}_\k-\delta)+\Omega R}, 
\end{align}
where we have introduced the parameter $R$
\begin{align} \nonumber
R= \frac{g_{23} \sum_\q \varphi_\q s_\q}{g_{13}\sum_\q \varphi_\q c_\q}. 
\end{align}

Multiplying \eqref{eq:Schrodingera} by $g_{13}$ and \eqref{eq:Schrodingerb} by  $g_{23}$, dividing both sides by $(E-...)$ and summing over $\k$, these equations can be rearranged as 
\begin{subequations}
\begin{align} 
\frac{1}{g_{13}}
&=\sum_{\k}\frac{1}{E -\bar{\epsilon}_\k-\frac{\Omega}{2} t_\k} ,\\
\frac{1}{g_{23}}
&=\sum_{\k}\frac{1}{E -\bar{\epsilon}_\k-\delta-\frac{\Omega}{2 t_\k} }  .
\end{align}
\end{subequations}
Using Eq.~\eqref{eq:tk13232body}, we obtain
\begin{subequations}
\begin{align} 
\frac{1}{g_{13}}
=\sum_{\k}\frac{E-\bar{\epsilon}_\k-\delta+\frac{\Omega}{2}R }{(E-\bar{\epsilon}_\k)(E-\bar{\epsilon}_\k-\delta)-\frac{\Omega^2}{4}} ,\\
\frac{1}{g_{23}}
=\sum_{\k}\frac{E-\bar{\epsilon}_\k+\frac{\Omega}{2R} }{(E-\bar{\epsilon}_\k)(E-\bar{\epsilon}_\k-\delta)-\frac{\Omega^2}{4}}.
\end{align}
\end{subequations}
Here we can recognize
\begin{subequations} \label{eq:gpi}
\begin{align} 
\frac{1}{g_{13}}
&=\Pi_{11}(E)+R\Pi_{12}(E) ,\\
\frac{1}{g_{23}}
&=\Pi_{22}(E)+\Pi_{21}(E)/R ,
\end{align}
\end{subequations}
with the functions $\Pi_{ij}$ defined in Eq. \eqref{eq.Piij}.
Isolating $R$ in the first line and injecting it in the second we obtain
\begin{align} \label{eq:det}
\det[\mathbf{g}^{-1}-\mathbf{\Pi}(E)]=0,
\end{align}
which corresponds to the equation for the poles of the dressed $T$ matrix \eqref{eq:tmat2by2SM}.

The solutions of \eqref{eq:det} correspond to bound states with energy $E=\epsilon_{-}-\varepsilon$ (where the binding energy $\varepsilon>0$ is measured from $\epsilon_{-}$).
One can then use \eqref{eq:gpi} to determine $R$
\begin{align} \label{eq:R}
R(\varepsilon)
&=\frac{\frac{1}{g_{13}}-\Pi_{11}(\epsilon_{-}-\varepsilon)}{\Pi_{12}(\epsilon_{-}-\varepsilon)},
\\ \nonumber
&=\frac{\frac{1}{a_{13}}-\left( c^2 \sqrt{2 m_r\varepsilon}+s^2 \sqrt{2 m_r(\varepsilon+\epsilon_{+}-\epsilon_{-})}\right)}{cs\left(\sqrt{2 m_r(\varepsilon+\epsilon_{+}-\epsilon_{-})}- \sqrt{2 m_r\varepsilon}\right)}.
\end{align}

We can also find the wave function of the bound state in a manner analogous to that used in Ref.~\cite{Liprb2021}.  Since the right hand side of the Schr\"{o}dinger equation \eqref{eq:Schrodinger} is independent of momentum, we must have 
 \begin{subequations}  \label{eq:WFk}
\begin{align} \nonumber
 \varphi_\k c_\k &=\frac{\mathcal{C}}{\varepsilon-\epsilon_{-} +\bar{\epsilon}_\k+\frac{\Omega}{2} t_\k} \\
  &=\mathcal{C}\left(\frac{c^2-cs R}{\varepsilon+\bar{\epsilon}_\k} +\frac{s^2+cs R}{\varepsilon+\epsilon_{+}-\epsilon_{-}+\bar{\epsilon}_\k}\right)  , 
\\ \nonumber  \varphi_\k s_\k &=\frac{\mathcal{S}}{\varepsilon-\epsilon_{-} +\bar{\epsilon}_\k+\delta +\frac{\Omega}{2 t_\k} } 
\\
  &=\mathcal{S}\left(\frac{s^2-cs/R}{\varepsilon+\bar{\epsilon}_\k} +\frac{c^2+cs/R}{\varepsilon+\epsilon_{+}-\epsilon_{-}+\bar{\epsilon}_\k}\right) ,
\end{align}
 \end{subequations}
where the coefficients $\mathcal{C}$ and $\mathcal{S}$ are determined by the normalization condition of the wave function and Eq.~\eqref{eq:tk13232body}. 
From \eqref{eq:tk13232body}, we have the relation $ \mathcal{S}/\mathcal{C}=R$, with the sign of $R$ determined by Eq.~\eqref{eq:R}, while the normalization imposes
\begin{align} \nonumber
\frac{1}{\mathcal{C}^2}&= \sum_\k\frac{1}{(\varepsilon-\epsilon_{-}  +\bar{\epsilon}_\k+\frac{\Omega}{2} t_\k)^2}+ \frac{R^2}{(\varepsilon-\epsilon_{-}  +\bar{\epsilon}_\k+\delta +\frac{\Omega}{2 t_\k})^2} \\
&=\frac{m_r^{3/2}}{2\sqrt{2}\pi}\left[  \frac{\left( c R+s \right)^2}{\sqrt{\varepsilon+\epsilon_{+}-\epsilon_{-} }}+  \frac{\left( s R -c\right)^2}{\sqrt{\varepsilon}} \right].
\end{align}

In position space, the Fourier transforms of \eqref{eq:WFk}, i.e., the two components entering the wave function in Eq.~\eqref{eq:2bodywavefunction}, take the forms 
 \begin{subequations}  \label{eq:WFr}
\begin{align} 
\varPhi_1(r) &\propto \frac{\mathcal{C}}{r} \left[e^{-r/l_{-}}(c^2-csR)+e^{-r/l_{+}}(s^2+csR)\right], 
\\ 
\varPhi_2(r) &\propto \frac{\mathcal{S}}{r} \left[e^{-r/l_{-}}(s^2-cs/R)+e^{-r/l_{+}}(c^2+cs/R)\right],
\end{align}
 \end{subequations}
respectively, where $l_{-}=1/\sqrt{2m_r\varepsilon}$ and $l_{+}=1/\sqrt{2m_r(\varepsilon+\epsilon_{+}-\epsilon_{-})}$. Interestingly, we can observe that the new hybrid bound states exhibit two characteristic localization lengths $l_{\pm}$, which is due to the presence of  the two scattering continua for lower and upper dressed particles. We also note that the two-component nature of the hybrid wave function \eqref{eq:WFr} gives rise to a radially dependent pseudospin texture.

\begin{figure}[tbp] 
    \includegraphics[width=0.5\linewidth]{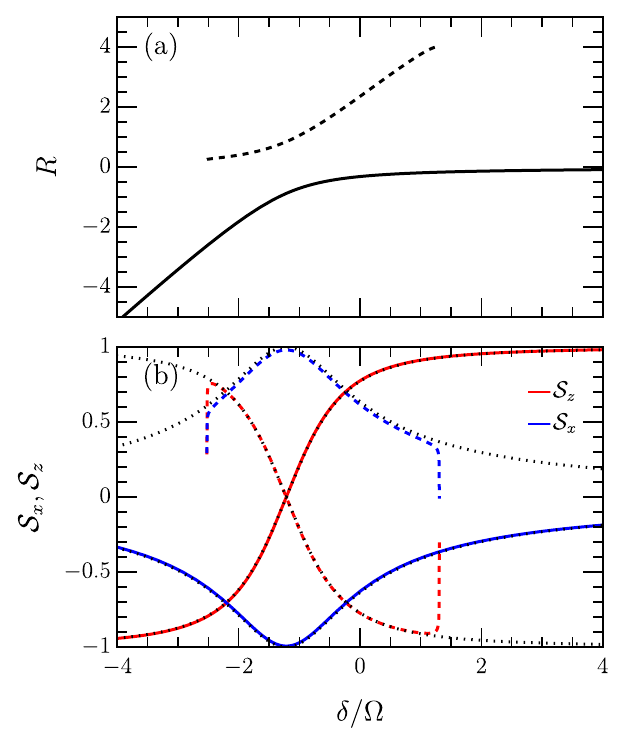}
\caption{Properties of the hybrid bound states. (a) Ratio $R$ versus detuning for the ground (solid line) and excited states (dashed line). (b) Magnetization vector component $\mathcal{S}_z$ (red) and $\mathcal{S}_x$ (blue) for the ground (solid line) and excited states (dashed line). The dotted-black lines show the magnetization obtained from the effective Hamiltonian \eqref{eq:Heff}. The bare scattering lengths are $(a_{13},a_{23})\sqrt{2m_r\Omega}=(0.6,0.8)$ as in Fig.~3(c) of the main text.}
\label{fig:BSWF}
\end{figure}

The knowledge of the wavefunction allows us to access some properties of the hybrid bound states as shown in Figure \ref{fig:BSWF} for the set of scattering lengths used in Fig.~3(c) of the main text.
Panel (a) shows the ratios $R$ \eqref{eq:R} and we see that they have opposite signs for the excited and the ground states. This tells us that the two hybrid bound states are made up of bare dimers that are in- and out-of-phase, respectively.
In panel (b), we plot the magnetization vector components obtained from the bound-state wavefunction using 
 \begin{subequations}
  \begin{align} \label{eq:magnetizationdef}
\mathcal{S}_z&=\frac{\sum_\k\bra{\psi} \hat{f}_{\k1}^{\dagger}\hat{f}_{\k1}-\hat{f}_{\k2}^{\dagger}\hat{f}_{\k2}\ket{\psi}}{\sum_\k\bra{\psi} \hat{f}_{\k1}^{\dagger}\hat{f}_{\k1}+\hat{f}_{\k2}^{\dagger}\hat{f}_{\k2}\ket{\psi}},\\
\mathcal{S}_x&=\frac{\sum_\k\bra{\psi} \hat{f}_{\k1}^{\dagger}\hat{f}_{\k2}+\hat{f}_{\k2}^{\dagger}\hat{f}_{\k1}\ket{\psi}}{\sum_\k\bra{\psi} \hat{f}_{\k1}^{\dagger}\hat{f}_{\k1}+\hat{f}_{\k2}^{\dagger}\hat{f}_{\k2}\ket{\psi}} .
\end{align}
      \end{subequations}
These magnetization components expose the differing nature of the two hybrid bound states. Indeed, we see  that the extrema of $\mathcal{S}_x$ have opposite signs and that the slope of $\mathcal{S}_z$ at its zero-crossing is opposite for the excited and ground states.  
In addition, the dotted lines in Fig.~\ref{fig:BSWF}(b) show the magnetization obtained using the eigenvectors of the the following effective 2 by 2 Hamiltonian 
  \begin{align} \label{eq:Heff}
H_{\text{eff}}&=\begin{pmatrix}
    -\varepsilon_{13} & \frac{\Omega}{2} \\  \frac{\Omega}{2}& -\varepsilon_{23}+\delta \end{pmatrix} ,
\end{align}
whose eigenvalues correspond to Eq. (13) of the main text.
We see that the magnetization obtained from the effective model is very accurate for the ground state while it is only accurate at small detunings for the excited state.  Indeed, the effective model \eqref{eq:Heff} does not account for the presence of the scattering continuum and its results are only reasonable when the bound states are well separated from it.

\end{document}